\documentclass{article}
\usepackage{spconf,amsmath,graphicx}
\usepackage{cite}

\title{ADVERSARIAL ATTACKS ON GMM i-vector BASED SPEAKER VERIFICATION SYSTEMS}
\name{Xu Li$^{\star}$, Jinghua Zhong$^{\star,\dagger}$, Xixin Wu$^{\star}$, Jianwei Yu$^{\star}$, Xunying Liu$^{\star}$, Helen Meng$^{\star}$}
\address{$^{\star}$Department of Systems Engineering and Engineering Management, \\ The Chinese University of Hong Kong\\
$^{\dagger}$SpeechX Limited, Shenzhen, China\\
\begin{footnotesize}
\texttt{\{xuli, jhzhong, wuxx, jwyu, xyliu, hmmeng\}@se.cuhk.edu.hk}
\end{footnotesize}
}

\begin{document}
\ninept
%
\maketitle
\begin{abstract}
This work investigates the vulnerability of Gaussian Mixture Model (GMM) i-vector based speaker verification systems to adversarial attacks, and the transferability of adversarial samples crafted from GMM i-vector based systems to x-vector based systems. 
In detail, we formulate the GMM i-vector system as a scoring function of enrollment and testing utterance pairs. Then we leverage the fast gradient sign method (FGSM) to optimize testing utterances for adversarial samples generation.
These adversarial samples are used to attack both GMM i-vector and x-vector systems. We measure the system vulnerability by the degradation of equal error rate and false acceptance rate. Experiment results show that GMM i-vector systems are seriously vulnerable to adversarial attacks, and the crafted adversarial samples prove to be transferable and pose threats to neural network speaker embedding based systems (e.g. x-vector systems).
\end{abstract}

\section{Introduction}
\label{sec:intro}

Automatic speaker verification (ASV) systems aim at confirming a spoken utterance against a speaker identity claim. Through decades of development, the speaker verification community has made great progress and applied this technology in many biometric authentication cases, such as voice activation in electronic devices, ebanking authentication, etc. 

However, past research has shown that ASV systems are vulnerable to malicious attacks via spoofing speech, such as impersonation \cite{zetterholm2004comparison,wu2015spoofing}, replay \cite{wu2015spoofing,wu2014study}, speech synthesis \cite{shchemelinin2013examining,shchemelinin2014vulnerability} and voice conversion  \cite{kinnunen2012vulnerability}. Among these four types of spoofing attack, replay, speech synthesis and voice conversion pose the most serious threats to ASV systems. The created spoofing speech could sound extremely close to the voice of the target person.
Moreover, ASV systems could also be spoofed even though the spoofing speech sounds like the voice of the imposter from human perception. It will expose the systems to some other dangerous situations, such as controlling one's voice-operated devices in place of the real owner unbeknownst to him/her. These threats could be posed by adversarial attacks.

According to \cite{szegedy2013intriguing,ebrahimi2018adversarial,goodfellow2014explaining}, deep neural networks (DNNs) with impressive performance can be vulnerable to simple adversarial attacks in many tasks, such as face recognition \cite{agarwal2018image,sharif2016accessorize}, image classification \cite{liu2016delving,athalye2017synthesizing}, and speech recognition \cite{carlini2018audio}. However, to the best of our knowledge, the only work that applied adversarial attack into ASV systems is \cite{kreuk2018fooling} where they verified the vulnerability of an end-to-end ASV system to adversarial attacks. Briefly, there are three representative ASV frameworks: i-vector speaker embedding based systems \cite{dehak2010front,kenny2012small,prince2007probabilistic,garcia2011analysis}, neural network (NN) speaker embedding based systems \cite{variani2014deep,snyder2018x} and end-to-end approaches \cite{zhang2016end,heigold2016end,snyder2016deep}. While end-to-end based systems have proved to be vulnerable to adversarial attacks, the robustness of other approaches, including GMM i-vector based systems and NN speaker embedding based systems (e.g. x-vector systems we implement), still remains to be explored. GMM i-vector based systems are widely applied to biometric authentication, and it is imperative to investigate their robustness to such adversarial attacks.

Adversarial attacks aim at perturbing the system input in a purposefully designed way to make the system behave incorrectly. The perturbations are usually subtle so that human cannot perceive differences between the adversarial and original inputs. There are two main attack scenarios: white box attack and black box attack. The white box attack allows the attacker to access the complete parameters of the system, so that the system function could be directly involved to optimize the input perturbations. The black box attacker only has the access to the system's input and output, and the adversarial samples are usually crafted by other substitute systems.

For adversarial samples generation, many algorithms were previously proposed to solve the perturbation optimization problem, such as fast gradient sign method (FGSM) \cite{goodfellow2014explaining}, basic iterative method (BIM) \cite{kurakin2016adversarial} and DeepFool \cite{moosavi2016deepfool}. In this work, we simply adopt FGSM to verify the ASV system's vulnerability to adversarial attacks.

This work focuses on the vulnerability of GMM i-vector systems to adversarial attacks, and the transferability of adversarial samples crafted from i-vector systems to x-vector systems. Specifically, we perform both white and black box attacks on GMM i-vector systems and black box attacks on x-vector systems. The detailed attack configurations are illustrated in Section~\ref{sec:adv-examples-gen}. In this work, FGSM \cite{goodfellow2014explaining} is adopted for adversarial perturbations optimization. Our codes have been made open-source\footnote{https://github.com/lixucuhk/adversarial-attack-on-GMM-i-vector-based-speaker-verification-systems}.

This paper is organized as follows. Section~\ref{sec:asv-systems} introduces the ASV systems adopted in our experiments. The adversarial attack configurations and the FGSM optimization algorithm are illustrated in Section~\ref{sec:adv-examples-gen}. Section~\ref{sec:exp-setup} and \ref{sec:experiments} describe the experimental setup and results, respectively. Finally, Section~\ref{sec:conclusion} concludes this work.

\section{automatic speaker verification systems}
\label{sec:asv-systems}

This work includes the GMM i-vector and x-vector systems in the experiments. Both kinds of systems have two parts: a front-end for utterance-level speaker embedding extraction and a back-end for speaker similarity scoring. The probabilistic linear discriminant analysis (PLDA) back-end is adopted in all the experiments.

\subsection{Gaussian Mixture Model i-vector extraction}
\begin{figure}[t]
    \centering
    \includegraphics[width=0.5\textwidth]{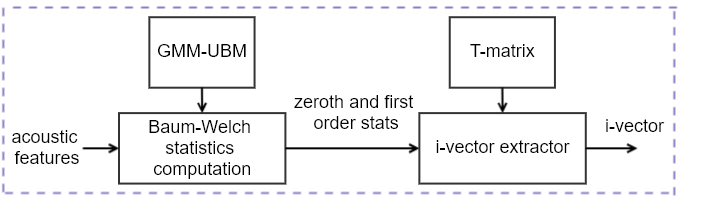}
    \caption{The illustration of GMM i-vector extractor}
    \label{fig:i-vector-extractor}
\end{figure}

The illustration of GMM i-vector extractor \cite{dehak2010front} is shown in Fig.~\ref{fig:i-vector-extractor}. It consists of a Gaussian Mixture Model-universal background model (GMM-UBM) and a total variability matrix ($\boldsymbol{T}$ matrix). Given the acoustic features of utterance $i$, GMM-UBM is used to extract the zeroth ($\boldsymbol{N_i}$) and first ($\boldsymbol{\tilde{f_i}}$) order statistics by Baum-Welch statistics computation. The statistics information is incorporated with $\boldsymbol{T}$ matrix to extract i-vector $\boldsymbol{\omega_i}$ as Eq.~\ref{eq:i-vector-extraction},

\begin{equation}
    \boldsymbol{\omega_i} = \boldsymbol{L_i}^{-1}\boldsymbol{T}^\top(\boldsymbol{\Sigma^{(b)}})^{-1}\boldsymbol{\tilde{f_i}}
    \label{eq:i-vector-extraction}
\end{equation}
where $\boldsymbol{L_i} = \boldsymbol{I}+\boldsymbol{T}^\top(\boldsymbol{\Sigma^{(b)}})^{-1}\boldsymbol{N_i}\boldsymbol{T}$, $\boldsymbol{I}$ is the identity matrix and $\boldsymbol{\Sigma^{(b)}}$ is the covariance matrix of GMM-UBM.

\subsection{x-vector extraction}
The x-vector extractor \cite{snyder2018x} leverages DNNs to produce speaker-discriminative embeddings. It consists of frame- and utterance-level extractors. At the frame level, acoustic features are fed forward by several layers of time delay neural network (TDNN). At the utterance level, statistics pooling layer aggregates all the last frame-level layer's outputs and computes their mean and standard deviation. The mean and standard deviation are concatenated together and propagated through utterance-level layers and finally softmax output layer. In the testing stage, given acoustic features of an utterance, the embedding layer output is extracted as the x-vector.


\subsection{Probabilistic linear discriminant analysis back-end}
PLDA is a supervised version of factor analysis \cite{fruchter1954introduction}. It models i-vectors/x-vectors ($\boldsymbol{\omega}$) by Eq.~\ref{eq:plda},

\begin{equation}
    \boldsymbol{\omega} = \boldsymbol{m} + \boldsymbol{\Phi}\boldsymbol{\beta} +\boldsymbol{\epsilon_r}
    \label{eq:plda}
\end{equation}
where $\boldsymbol{m}$ is a global bias term, the columns of $\boldsymbol{\Phi}$ provides a basis of speaker-specific subspace, and $\boldsymbol{\beta} \in N(\boldsymbol{0},\boldsymbol{I})$ is a latent speaker-identity vector. The residual term $\boldsymbol{\epsilon_r} \in N(\boldsymbol{0}, \boldsymbol{\Sigma})$ has a Gaussian distribution with zero mean and a full corvariance matrix $\boldsymbol{\Sigma}$. The model parameters \{$\boldsymbol{m}$, $\boldsymbol{\Phi}$, $\boldsymbol{\Sigma}$\} are estimated with the expectation-maximization (EM) algorithm on the training set.

In the testing stage, score $S$ is estimated as a log likelihood ratio of two conditional probabilities (Eq.~\ref{eq:plda-score}). $\boldsymbol{\omega_1}$ and $\boldsymbol{\omega_2}$ are i-vectors/x-vectors extracted from enrollment and testing utterances, respectively. $\mathcal{H}_s$ is the hypothesis that two utterances belong to the same identity, whereas $\mathcal{H}_d$ is the opposite.
\begin{align}
    \nonumber
    S =& \log \frac{P(\boldsymbol{\omega_1},\boldsymbol{\omega_2}|\mathcal{H}_s)}{P(\boldsymbol{\omega_1},\boldsymbol{\omega_2}|\mathcal{H}_d)} \\
          =& \log \frac{P(\boldsymbol{\omega_1},\boldsymbol{\omega_2}|\mathcal{H}_s)}{P(\boldsymbol{\omega_1}|\mathcal{H}_d)P(\boldsymbol{\omega_2}|\mathcal{H}_d)}
    \label{eq:plda-score}
\end{align}

\section{adversarial samples generation}
\label{sec:adv-examples-gen}

\begin{figure}[t]
    \centering
    \includegraphics[width=0.5\textwidth]{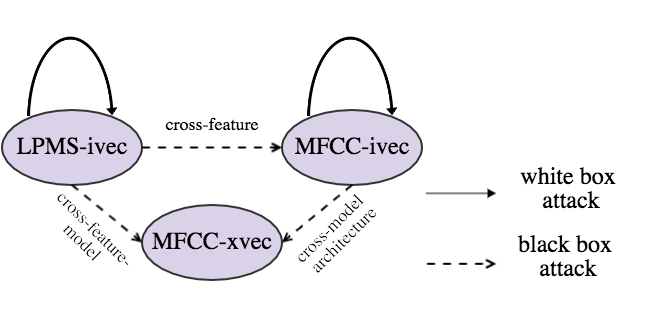}
    \caption{The adversarial attack configuration. Each arrow represents an attack setting, where it points from the source model to the target model.}
    \label{fig:adv-attack-config}
\end{figure}

In this work, we investigate the vulnerability of ASV systems to adversarial attacks, including the white box attack and black box attacks in terms of cross-feature, cross-model architecture and cross-feature-model settings. Three ASV models are well-trained in our experiments: Mel-frequency cepstral coefficient (MFCC) based GMM i-vector system (MFCC-ivec), log power magnitude spectrum (LPMS) based GMM i-vector system (LPMS-ivec) and MFCC based x-vector system (MFCC-xvec). The white box attack and black box mutual attack settings are illustrated in Fig.~\ref{fig:adv-attack-config}. Two white box attacks are performed on MFCC-ivec and LPMS-ivec systems, respectively. The three black box attack settings are designed as follows: LPMS-ivec attacks MFCC-ivec (cross-feature), MFCC-ivec attacks MFCC-xvec (cross-model architecture) and LPMS-ivec attacks MFCC-xvec (cross-feature-model). The last two settings are to investigate the transferability of adversarial samples crafted from GMM i-vector systems to x-vector systems.

To perform the attacks above, adversarial samples need to be crafted from the source models, i.e. LPMS-ivec and MFCC-ivec in our case. In this work, we generate adversarial samples at the acoustic feature level, i.e. either MFCC or LPMS. In the black box attacks of cross-feature and cross-feature-model settings, the generated adversarial features (LPMS) are first inverted to audios, then the acoustic features adopted in the target models are extracted from the audios to perform attacks. In other cases, the adversarial features could be directly used for attacks.


\subsection{Fast gradient sign method}
In a general case, we denote a system function as $f$ with parameters $\theta$. Given the original input $\boldsymbol{x}$, we search for an adversarial perturbation $\boldsymbol{\delta_x}$ to be added to $\boldsymbol{x}$ to generate the adversarial sample $\boldsymbol{x^\star} = \boldsymbol{x} + \boldsymbol{\delta_x}$. This adversarial perturbation $\boldsymbol{\delta_x}$ is optimized by maximizing the deviation between the system's prediction given adversarial input $\boldsymbol{x^\star}$ and the ground truth $y$, as shown in Eq.~\ref{eq:formal-adv},
\begin{equation}
    \boldsymbol{\delta_x} = \arg\max_{\Vert \boldsymbol{\delta_x} \Vert_p \leq \epsilon} L(f_{\theta}(\boldsymbol{x}+\boldsymbol{\delta_x}), y)
    \label{eq:formal-adv}
\end{equation}
The deviation is measured by a loss function $L$ between the system prediction and the ground truth, which is usually adopted as either cross entropy for classification tasks or mean square error for regression tasks. The $p$-norm constraint on $\boldsymbol{\delta_x}$ guarantees that human cannot distinguish the adversarial sample $\boldsymbol{x^\star}$ from the original $\boldsymbol{x}$, and the upper bound $\epsilon$ represents the perturbation degree. Note that the system parameters $\theta$ are always fixed during the optimization procedure, which means adversarial attackers only revise the input to attack systems instead of the system parameters.

In this work, we adopt fast gradient sign method (FGSM) \cite{goodfellow2014explaining} to solve the optimization problem in Eq.~\ref{eq:formal-adv}. It perturbs the original input $\boldsymbol{x}$ towards the gradient of the loss function $L$ w.r.t. $\boldsymbol{x}$ to generate adversarial samples. Specifically, it specializes the norm $p$ in Eq.~\ref{eq:formal-adv} as $\infty$, and gives the solution as Eq.~\ref{eq:fsgm-adv},
\begin{equation}
    \boldsymbol{\delta_x} = \epsilon \times sign(\nabla_{\boldsymbol{x}} L(f_{\theta}(\boldsymbol{x}), y))
    \label{eq:fsgm-adv}
\end{equation}
where the function $sign(\cdot)$ takes the sign of the gradient, and $\epsilon$ represents the perturbation degree.


\subsection{Problem formulation}

Adversarial samples are generated from GMM i-vector systems (LPMS-ivec and MFCC-ivec systems). To formulate it into an optimization problem, we denote the GMM i-vector system by two functions: i-vector extractor $f$ with parameters $\theta_1$ and PLDA scoring function $S$ with parameters $\theta_2$. Normalization steps, such as i-vector normalization, are implicitly included within the functions.

The acoustic features of each utterance $i$ are denoted as matrix $\boldsymbol{X_i}$, where each column represents the acoustic feature of each frame and the horizontal axis represents the time sequence. Given the acoustic features of two utterances $\boldsymbol{X_i}$ and $\boldsymbol{X_j}$, the corresponding i-vectors are derived as $\boldsymbol{\omega_i}=f_{\theta_1}(\boldsymbol{X_i})$ and $\boldsymbol{\omega_j}=f_{\theta_1}(\boldsymbol{X_j})$, respectively. Then the final similarity score (PLDA score adopted in our experiments) based on these two i-vectors is derived as $S_{\theta_2}(f_{\theta_1}(\boldsymbol{X_i}), f_{\theta_1}(\boldsymbol{X_j}))$.

In the ASV testing stage, each testing trial consists of one enrollment utterance $\boldsymbol{X_i}$ and one testing utterance $\boldsymbol{X_j}$. According to the speaker identities behind these two utterances, there could be two kinds of trials: target trials and non-target trials. In target trials, the speaker identities behind the enrollment and testing utterances belong to one person. In non-target trials, the identities belong to two different persons. To mimic a realistic model attack, we keep the enrollment features $\boldsymbol{X_i}$ and the system parameters $\{\theta_1, \theta_2\}$ fixed and revise the testing features $\boldsymbol{X_j}$ for adversarial samples generation. For target trials, we search perturbations $\boldsymbol{\delta_X}$ to be added to $\boldsymbol{X_j}$ to minimize the similarity score. For non-target trials, we search perturbations to maximize the score. This could lead the ASV systems to make wrong decisions for both target and non-target trials. The optimization problem formulation is shown in Eq.~\ref{eq:our-case-0}, and the solution given by FGSM is derived as Eq.~\ref{eq:our-case-1} and \ref{eq:our-case-2}.

\begin{align}
    &\boldsymbol{\delta_X} = \arg \max_{\Vert \boldsymbol{\delta_X} \Vert_p \leq \epsilon} k \times S_{\theta_2}(f_{\theta_1}(\boldsymbol{X_i}), f_{\theta_1}(\boldsymbol{X_j}+\boldsymbol{\delta_X})) \label{eq:our-case-0} \\
    &\boldsymbol{\delta_X} = \epsilon \times k \times sign(\nabla_{\boldsymbol{X_j}} S_{\theta_2}(f_{\theta_1}(\boldsymbol{X_i}), f_{\theta_1}(\boldsymbol{X_j}))) \label{eq:our-case-1} \\
    &k = \{ \begin{array}{cc}
     -1, & \text{target trial} \\
     1, & \text{non-target trial}
    \end{array}
    \label{eq:our-case-2}
\end{align}

\section{Experimental Setup}
\label{sec:exp-setup}

The dataset used in this experiment is Voxceleb1 \cite{nagrani2017voxceleb}, which consists of short clips of human speech. There are in total 148,642 utterances for 1251 speakers. Consistent with \cite{nagrani2017voxceleb}, 4874 utterances for 40 speakers are reserved for testing, to generate trials and perform adversarial attacks. The remaining utterances are used for training our SV models. In addition, we apply data augmentation \cite{snyder2018x} when training x-vector embedding networks.

Mel-frequency cepstral coefficients (MFCCs) and log power magnitude spectrums (LPMSs) are adopted as acoustic features in this experiment. To extract MFCCs, a pre-emphasis with coefficient of 0.97 is adopted. ``Hamming" window having size of 25ms and step-size of 10ms is applied to extract a frame, and finally 24 cepstral coefficients are kept. For LPMSs, ``blackman" window having size of 8ms and step-size of 4ms is adopted. No pre-emphasis is applied. 

\subsection{ASV model configuration}
In the GMM i-vector system setup, only voice activity detection (VAD) is applied for preprocessing acoustic features. 2048-mixture UBM with full covariance matrix cooperates $\boldsymbol{T}$ matirx with a 400-dimension i-vector space. The i-vectors are centered and length-normalized before PLDA modeling.

In the x-vector system, cepstral mean and variance normalization (CMVN) and VAD are adopted to preprocess the acoustic features. The setup of x-vector embedding network is commonly consistent with \cite{snyder2018x}. Extracted x-vectors are centered and projected using a 200-dimension LDA, then length-normalized before PLDA modeling.



\begin{table*}
\centering
\caption{EER (\%) of the target systems under black box attack with different perturbation degrees ($\epsilon$).}
\begin{tabular}{c|c|c|c|c|c|c|c|c}
\hline
 & $\epsilon = 0$ & $\epsilon = 0.3$ & $\epsilon = 1$ & $\epsilon = 5$ & $\epsilon = 10$ & $\epsilon = 20$ & $\epsilon = 30$ & $\epsilon = 50$ \\
\hline
LPMS-ivec attacks MFCC-ivec  & 7.20 & 8.83 & 13.82 & 50.02 & 69.04 & \textbf{74.62} & 74.59 & 63.24 \\
\hline
MFCC-ivec attacks MFCC-xvec & 6.62 & 8.52 & 14.06 & 57.43 & \textbf{74.32} & 60.85 & 54.07 & 51.34 \\
\hline
LPMS-ivec attacks MFCC-xvec & 6.62 & 7.42 & 9.49 & 25.47 & 37.51 & 43.89 & \textbf{48.48} & 48.39 \\
\hline
\end{tabular}
\label{tab:black-box-attack-eer}
\end{table*}

\begin{table}[t]
\centering
\caption{FAR (\%) of the GMM i-vector systems under white box attack with different perturbation degrees ($\epsilon$).}
\begin{tabular}{c|c|c|c|c|c}
\hline
 & $\epsilon = 0$ & $\epsilon = 0.3$ & $\epsilon = 1$ & $\epsilon = 5$ & $\epsilon = 10$ \\
\hline
MFCC-ivec  & 7.20 & 82.91 & \textbf{96.87} & 18.14 & 16.65 \\
\hline
LPMS-ivec & 10.24 & 96.78 & \textbf{99.99} & 99.64 & 69.95 \\
\hline
\end{tabular}
\label{tab:white-box-attack-far}
\end{table}

\begin{table}[t]
\centering
\caption{EER (\%) of the GMM i-vector systems under white box attack with different perturbation degrees ($\epsilon$).}
\begin{tabular}{c|c|c|c|c|c}
\hline
 & $\epsilon = 0$ & $\epsilon = 0.3$ & $\epsilon = 1$ & $\epsilon = 5$ & $\epsilon = 10$ \\
\hline
MFCC-ivec  & 7.20 & 81.78 & \textbf{97.64} & 50.25 & 50.72 \\
\hline
LPMS-ivec & 10.24 & 94.04 & \textbf{99.95} & 99.77 & 88.6 \\
\hline
\end{tabular}
\label{tab:white-box-attack-eer}
\end{table}


\subsection{Evaluation metrics}
The evaluation metrics of ASV systems could be false rejection rate (FRR), false acceptance rate (FAR) and equal error rate (EER). The FRR and FAR measure the classification error for target and non-target trials, respectively, and EER is the balanced metric where two error rates are equal. As realistic attack situations are mostly similar with non-target trials, we are more concerned with the FAR increase of the system after adversarial attacks. Actually for target trials, instead of adversarial perturbations, simple random noise added to the input could also cause the system to fail in recognizing the owner's voice, so the FRR increase could not reflect the vulnerability of ASV systems to adversarial attacks well. With the considerations above, we measure the system vulnerability to adversarial attacks by the increase of FAR and EER.

\subsection{ABX test}
To evaluate the auditory indistinguishability of adversarial audios compared with the original audios, we perform the ABX test \cite{munson1950standardizing}, which is a forced choice test to identify detectable differences between two choices of sensory stimuli. The adversarial samples are generated from the LPMS-ivec by using FGSM with $\epsilon$ equal to 1. Each adversarial audio is reconstructed from the perturbed LPMSs and the phase of its corresponding original audio. In this work, 50 randomly selected original-adversarial audio (A and B) pairs are presented to the listeners, and from each pair, one audio is chosen as the audio X. Eight listeners join this test, and they are asked to choose one audio from A and B, which is the audio X.

\section{experiment results}
\label{sec:experiments}

\begin{figure}[t]
    \centering
    \includegraphics[width=0.5\textwidth]{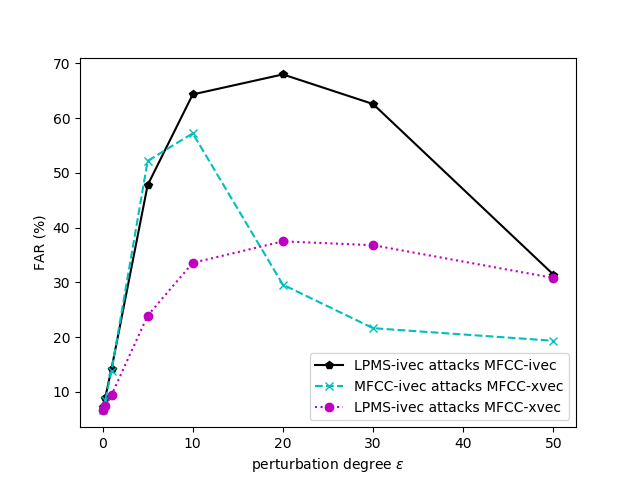}
    \caption{FAR (\%) of the target systems under black box attack with different perturbation degrees ($\epsilon$).}
    \label{fig:black-box-far}
\end{figure}

\subsection{ABX test results}
Experiment results show that the average accuracy of ABX test is \textbf{51.5\%}, which verifies that human cannot distinguish between the adversarial and corresponding original audios.

\subsection{White box attacks}
We perform the white box attack on both MFCC-ivec and LPMS-ivec systems. The FARs of the two systems under different perturbation degree $\epsilon$ are shown in Table~\ref{tab:white-box-attack-far}. Specifically, the columns where $\epsilon$ equals 0 exhibit the system performance without attack. It is observed that the FARs for both systems are dramatically increased by around 90\% when $\epsilon$ equals 1, where humans still cannot distinguish between the adversarial samples and the original ones. Similar results can be observed in the EER performance of the two systems, as shown in Table~\ref{tab:white-box-attack-eer}. This verifies that GMM i-vector systems are vulnerable to white box attacks. Moreover, from the columns where $\epsilon=5,10$ in Table~\ref{tab:white-box-attack-far} and \ref{tab:white-box-attack-eer}, we also observe that continuously increasing $\epsilon$ beyond some value can decrease attack effectiveness. One possible explanation is that continuously increasing $\epsilon$ makes the perturbations go beyond adversarial effectiveness and simply become noise. Noise cannot degrade ASV systems in terms of non-target trials, so the FARs begin to decrease. But the noise can still degrade ASV systems in terms of target trials, so the balanced metric EER tend to converge to around 50\%. This phenomenon is also observed in the experimental results of black box attacks.

\subsection{Black box attacks}
As configured in Section~\ref{sec:adv-examples-gen}, three black box attack settings are involved, i.e. LPMS-ivec attacks MFCC-ivec (cross-feature), MFCC-ivec  attacks MFCC-xvec (cross-model architecture) and LPMS-ivec attacks MFCC-xvec (cross-feature-model). The FAR fluctuations under these attack settings are illustrated in Fig.~\ref{fig:black-box-far}. It is observed that the FAR is dramatically increased in all three attack settings. Specifically, FAR increases the most by around 60\%, 50\% and 30\% in cross-feature, cross-model architecture and cross-feature-model settings, respectively. The most effective attack appears in the cross-feature setting, while the least effective attack appears in the cross-feature-model setting. This suggests that black box attack effectiveness can be affected by the gap between source and target system settings. The FAR degradation confirms that both GMM i-vector and x-vector systems are vulnerable to black box attacks. Specifically, the adversarial samples crafted from i-vector systems are transferable to x-vector systems. Similar results can be observed in the EER performance, as shown in Table~\ref{tab:black-box-attack-eer}. The phenomenon that x-vector systems are vulnerable to black box attacks also indicates their vulnerability to a more severe attack, i.e. white box attacks.

Moreover, we observe that a large $\epsilon$ value can still make an effective attack, e.g. 20 in the cross-feature setting. 
In this case, humans can perceive a small noise in the spoofing audios, but still confirm that the speaker identity does not change between the original and the corresponding spoofing audios.
Some adversarial samples and the corresponding system responses are illustrated in this URL\footnote{https://lixucuhk.github.io/adversarial-samples-attacking-ASV-systems}.

\section{conclusion}
\label{sec:conclusion}
In this work, we investigate the vulnerability of GMM i-vector systems to adversarial attacks and the adversarial transferability from i-vector systems to x-vector systems. Experimental results show that GMM i-vector systems are vulnerable to both white and black box attacks. The generated adversarial samples also prove to be transferable to NN speaker embedding based systems, e.g. x-vector systems. Further work will focus on protecting ASV systems against such adversarial attacks, e.g. involving adversarial training strategy to develop ASV systems.

\section{acknowledgement}
This work is partially supported by HKSAR Government's Research Grants Council General Research Fund (Project No. 14208718).

\bibliographystyle{IEEEbib}
\bibliography{strings,refs}

\begin{thebibliography}{10}

\bibitem{zetterholm2004comparison}
E.~Zetterholm, M.~Blomberg, and D.~Elenius,
\newblock ``A comparison between human perception and a speaker verification
  system score of a voice imitation,''
\newblock {\em evaluation}, vol. 119, pp. 116--4, 2004.

\bibitem{wu2015spoofing}
Z.~Wu, N.~Evans, T.~Kinnunen, J.~Yamagishi, F.~Alegre, and H.~Li,
\newblock ``Spoofing and countermeasures for speaker verification: A survey,''
\newblock {\em speech communication}, vol. 66, pp. 130--153, 2015.

\bibitem{wu2014study}
Z.~Wu, S.~Gao, E.~Cling, and H.~Li,
\newblock ``A study on replay attack and anti-spoofing for text-dependent
  speaker verification,''
\newblock in {\em Signal and Information Processing Association Annual Summit
  and Conference (APSIPA)}. IEEE, 2014, pp. 1--5.

\bibitem{shchemelinin2013examining}
V.~Shchemelinin and K.~Simonchik,
\newblock ``Examining vulnerability of voice verification systems to spoofing
  attacks by means of a {TTS} system,''
\newblock in {\em ICSC}. Springer, 2013, pp. 132--137.

\bibitem{shchemelinin2014vulnerability}
V.~Shchemelinin, M.~Topchina, and K.~Simonchik,
\newblock ``Vulnerability of voice verification systems to spoofing attacks by
  tts voices based on automatically labeled telephone speech,''
\newblock in {\em International Conference on Speech and Computer}. Springer,
  2014, pp. 475--481.

\bibitem{kinnunen2012vulnerability}
T.~Kinnunen, Z.~Wu, K.~Lee, F.~Sedlak, E.~Chng, and H.~Li,
\newblock ``Vulnerability of speaker verification systems against voice
  conversion spoofing attacks: The case of telephone speech,''
\newblock in {\em ICASSP}. IEEE, 2012, pp. 4401--4404.

\bibitem{szegedy2013intriguing}
C.~Szegedy, W.~Zaremba, I.~Sutskever, J.~Bruna, D.~Erhan, I.~Goodfellow, and
  R.~Fergus,
\newblock ``Intriguing properties of neural networks,''
\newblock {\em arXiv preprint arXiv:1312.6199}, 2013.

\bibitem{ebrahimi2018adversarial}
J.~Ebrahimi, D.~Lowd, and D.~Dou,
\newblock ``On adversarial examples for character-level neural machine
  translation,''
\newblock {\em arXiv preprint arXiv:1806.09030}, 2018.

\bibitem{goodfellow2014explaining}
I.~Goodfellow, J.~Shlens, and C.~Szegedy,
\newblock ``Explaining and harnessing adversarial examples,''
\newblock {\em arXiv preprint arXiv:1412.6572}, 2014.

\bibitem{agarwal2018image}
A.~Agarwal, R.~Singh, M.~Vatsa, and N.~Ratha,
\newblock ``Are image-agnostic universal adversarial perturbations for face
  recognition difficult to detect?,''
\newblock in {\em International Conference on Biometrics Theory, Applications
  and Systems (BTAS)}. IEEE, 2018, pp. 1--7.

\bibitem{sharif2016accessorize}
M.~Sharif, S.~Bhagavatula, L.~Bauer, and M.~K Reiter,
\newblock ``Accessorize to a crime: Real and stealthy attacks on
  state-of-the-art face recognition,''
\newblock in {\em ACM SIGSAC Conference on Computer and Communications
  Security}. ACM, 2016, pp. 1528--1540.

\bibitem{liu2016delving}
Y.~Liu, X.~Chen, C.~Liu, and D.~Song,
\newblock ``Delving into transferable adversarial examples and black-box
  attacks,''
\newblock {\em arXiv preprint arXiv:1611.02770}, 2016.

\bibitem{athalye2017synthesizing}
A.~Athalye, L.~Engstrom, A.~Ilyas, and K.~Kwok,
\newblock ``Synthesizing robust adversarial examples,''
\newblock {\em arXiv preprint arXiv:1707.07397}, 2017.

\bibitem{carlini2018audio}
N.~Carlini and D.~Wagner,
\newblock ``Audio adversarial examples: Targeted attacks on speech-to-text,''
\newblock in {\em IEEE Security and Privacy Workshops (SPW)}, 2018, pp. 1--7.

\bibitem{kreuk2018fooling}
F.~Kreuk, Y.~Adi, M.~Cisse, and J.~Keshet,
\newblock ``Fooling end-to-end speaker verification with adversarial
  examples,''
\newblock in {\em ICASSP}. IEEE, 2018, pp. 1962--1966.

\bibitem{dehak2010front}
N.~Dehak, P.~Kenny, R.~Dehak, P.~Dumouchel, and P.~Ouellet,
\newblock ``Front-end factor analysis for speaker verification,''
\newblock {\em IEEE Transactions on Audio, Speech, and Language Processing},
  vol. 19, no. 4, pp. 788--798, 2010.

\bibitem{kenny2012small}
P.~Kenny,
\newblock ``A small footprint i-vector extractor,''
\newblock in {\em The Speaker and Language Recognition Workshop}, 2012.

\bibitem{prince2007probabilistic}
S.~Prince and J.~Elder,
\newblock ``Probabilistic linear discriminant analysis for inferences about
  identity,''
\newblock in {\em International Conference on Computer Vision}. IEEE, 2007, pp.
  1--8.

\bibitem{garcia2011analysis}
D.~Garcia-Romero and C.~Espy-Wilson,
\newblock ``Analysis of i-vector length normalization in speaker recognition
  systems,''
\newblock in {\em Twelfth Annual Conference of the International Speech
  Communication Association}, 2011.

\bibitem{variani2014deep}
E.~Variani, X.~Lei, E.~McDermott, I.~Moreno, and J.~Gonzalez-Dominguez,
\newblock ``Deep neural networks for small footprint text-dependent speaker
  verification,''
\newblock in {\em ICASSP}. IEEE, 2014, pp. 4052--4056.

\bibitem{snyder2018x}
D.~Snyder, D.~Garcia-Romero, G.~Sell, D.~Povey, and S.~Khudanpur,
\newblock ``X-vectors: Robust {DNN} embeddings for speaker recognition,''
\newblock in {\em ICASSP}. IEEE, 2018, pp. 5329--5333.

\bibitem{zhang2016end}
S.~Zhang, Z.~Chen, Y.~Zhao, J.~Li, and Y.~Gong,
\newblock ``End-to-end attention based text-dependent speaker verification,''
\newblock in {\em Spoken Language Technology Workshop (SLT)}. IEEE, 2016, pp.
  171--178.

\bibitem{heigold2016end}
G.~Heigold, I.~Moreno, S.~Bengio, and N.~Shazeer,
\newblock ``End-to-end text-dependent speaker verification,''
\newblock in {\em ICASSP}. IEEE, 2016, pp. 5115--5119.

\bibitem{snyder2016deep}
D.~Snyder, P.~Ghahremani, D.~Povey, D.~Garcia-Romero, Y.~Carmiel, and
  S.~Khudanpur,
\newblock ``Deep neural network-based speaker embeddings for end-to-end speaker
  verification,''
\newblock in {\em Spoken Language Technology Workshop (SLT)}. IEEE, 2016, pp.
  165--170.

\bibitem{kurakin2016adversarial}
A.~Kurakin, I.~Goodfellow, and S.~Bengio,
\newblock ``Adversarial machine learning at scale,''
\newblock {\em arXiv preprint arXiv:1611.01236}, 2016.

\bibitem{moosavi2016deepfool}
S.~Moosavi-Dezfooli, A.~Fawzi, and P.~Frossard,
\newblock ``Deepfool: a simple and accurate method to fool deep neural
  networks,''
\newblock in {\em CVPR}, 2016, pp. 2574--2582.

\bibitem{fruchter1954introduction}
B.~Fruchter,
\newblock ``Introduction to factor analysis.,''
\newblock 1954.

\bibitem{nagrani2017voxceleb}
A.~Nagrani, J.~S. Chung, and A.~Zisserman,
\newblock ``Voxceleb: a large-scale speaker identification dataset,''
\newblock {\em arXiv preprint arXiv:1706.08612}, 2017.

\bibitem{munson1950standardizing}
W.~Munson and M.~Gardner,
\newblock ``Standardizing auditory tests,''
\newblock {\em The Journal of the Acoustical Society of America}, vol. 22, no.
  5, pp. 675--675, 1950.

\end{thebibliography}

\end{document}